\documentclass{article}
\usepackage[margin=1.0in]{geometry}                % See geometry.pdf to learn the layout options. There are lots.
\usepackage{graphicx}
\begin{document}
\large{
\vskip 6cm
\centerline{\bf A Note On Einstein, Bergmann, and the Fifth Dimension}
\vskip 1cm
\centerline{Edward Witten}
\centerline{Institute for Advanced Study}
\centerline{Princeton, NJ 08540 USA}

\bigskip

\begin{abstract}
This note is devoted to a detail concerning the work of Albert Einstein and Peter Bergmann on unified theories of electromagnetism and gravitation in 
five dimensions.  In their paper of 1938, Einstein and Bergmann were among the first to introduce the modern viewpoint in which a four-dimensional
theory that coincides with Einstein-Maxwell theory at long distances is derived from a five-dimensional theory with complete symmetry among all five dimensions.  
But then they drew back, modifying the theory in a way that spoiled the five-dimensional symmetry and looks contrived to modern readers.  Why?  According
to correspondence of Peter Bergmann with the author, the reason was that the more symmetric version of the theory predicts the existence of a new long range
field (a massless scalar field).  In 1938, Einstein and Bergmann did not wish to make this prediction.  (Based on a lecture at the Einstein Centennial Celebration
at the Library of Alexandria, June, 2005.)
\end{abstract}
\newpage
\bigskip
\bigskip

This note is devoted to a historical detail concerning the paper of Albert Einstein and Peter Bergmann, published in 1938, about unified theories of electromagnetism
and gravitation derived from five dimensions [1].  I read this paper for the first time over thirty years ago and immediately became curious about one point.
As I will explain shortly, Einstein and Bergmann began the paper by  introducing
a very modern point of view about a possible fifth dimension.  But  then they drew back, spoiling what a modern reader would see as the natural
beauty of the construction.  Why?

 My hunch was that this was because in 1938, Einstein and Bergmann did not wish to predict the existence of a new long range
field (a massless scalar field).  I wrote to Peter Bergmann inquiring about this point (fig. 1).  In my opinion, his response (fig. 2) confirmed my interpretation.  In this 
note, I will explain the issue and how I understand Bergmann's letter.

\begin{figure}
\begin{center}
\includegraphics[width=6.6in]{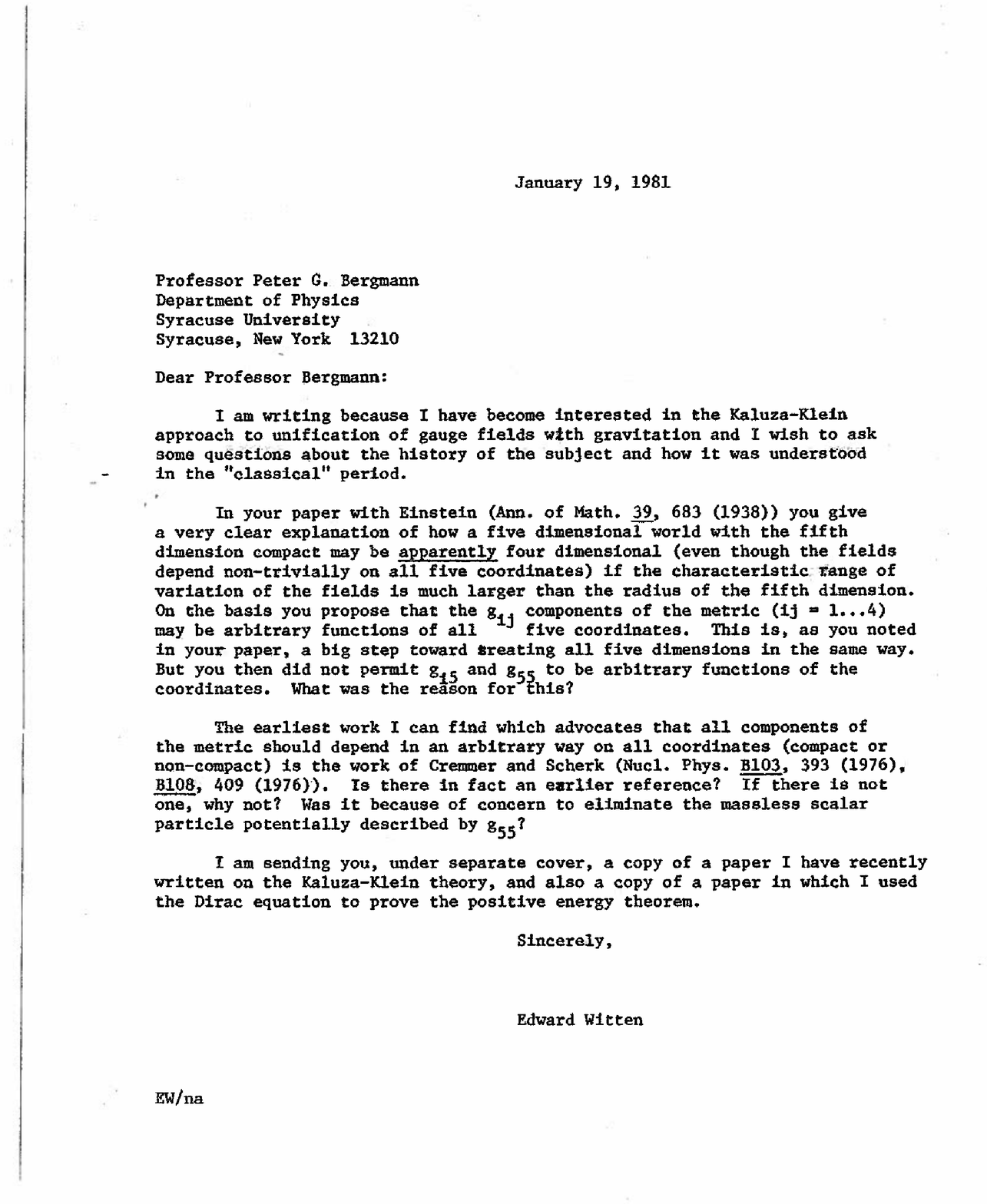}
\end{center}
\caption{
Letter of the author to Peter Bergmann.}
\end{figure}

\begin{figure}
\begin{center}
\includegraphics[width=6.6in]{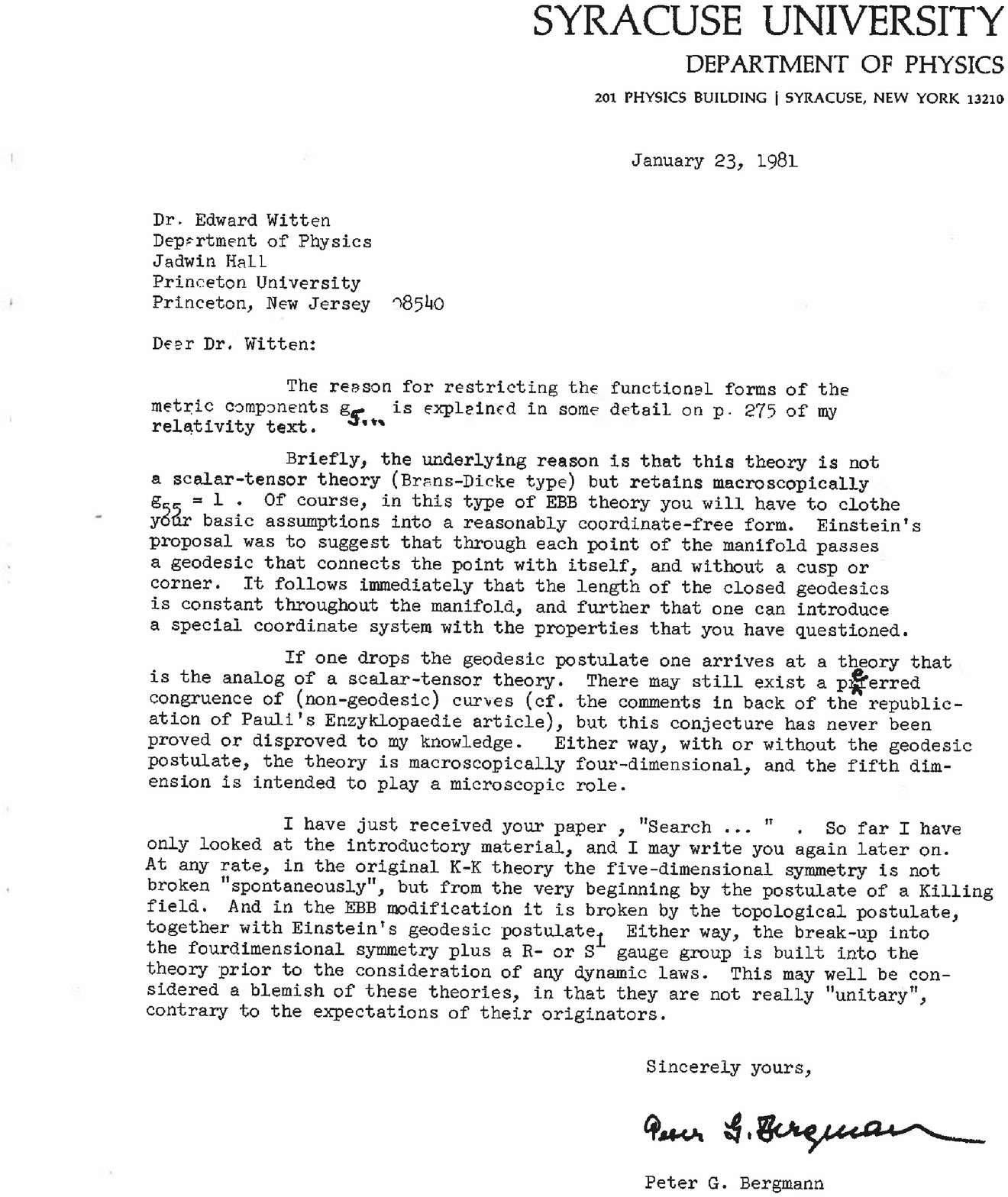}
\end{center}
\caption{
Peter Bergmann's response.}
\end{figure}

A major influence in Einstein's efforts to unify electromagnetism and gravitation was the 
 proposal made by Theodore Kaluza [2] around 1921, later independently discovered and extended by Oskar Klein [3] and commonly called Kaluza-Klein theory.
 In this proposal, in addition to the four dimensions of conventional relativity theory (three space dimensions and a fourth dimension of time) there is a fifth
 dimension; electromagnetism results from a gravitational field that is ``polarized" in the fifth dimension.\footnote{These original papers are 
 reprinted in English translation in reference [4], along with some other key papers, such as the one of Einstein and Bergmann that is our main interest here.  
 The book also contains a fascinating historical introduction to Kaluza-Klein theory with much little-known information.  For example, the first attempt at a 
 five-dimensional unification of electromagnetism with gravitation was actually made by Nordstrom, who in 1914 (before General Relativity!) attempted to 
 unify Maxwell's  theory with his own relativistic theory of gravitation, which is based on a massless scalar field that (in modern language) couples
 to the trace of the stress tensor.  Nordstrom's paper is reprinted in [4].}   

This idea feels very modern, and may well have some truth in it, though it is certainly not the whole story in the way that Einstein supposed.  
Actually, the original proposal by Kaluza was missing what to a modern reader is a very essential ingredient.  Kaluza introduced a fifth dimension
as a way to combine electromagnetism and gravitation, but he did not formulate field equations with five-dimensional symmetry.   Hence to a modern
reader, Kaluza did not have a symmetry between electromagnetism and gravitation and what he had was perhaps more a unified notation than a unified
theory.  (Klein was  closer to a modern viewpoint.  Einstein and Bergmann refer to Klein only vaguely and I will not attempt here to analyze the relation of the
work of Einstein and Bergmann to that of Klein.)  

The main novelty of Einstein and Bergmann was to take the fifth dimension seriously as a physical entity, not just an excuse to combine the
metric tensor and the electromagnetic potential as different components of a $5\times 5$ matrix.  In their introduction, they write:
 ``The theory presented here differs from Kaluza's in one essential point; we ascribe physical reality to the  fifth dimension whereas in Kaluza's theory this fifth dimension was introduced only in order to obtain new components of the metric tensor representing the electromagnetic field. Kaluza assumes the dependence of the field variables on the four coordinates $x^1,x^2,x^3,x^4$ and not on the fifth coordinate $x^0$ when a suitable coordinate system is 
chosen.   It is clear that this is due to the fact that the physical continuum is, according to our experience a four dimensional one.  We shall show, however, that it is possible to assign some meaning to the fifth coordinate without  contradicting the four dimensional character of the physical continuum."

To explain why the universe can appear to be four-dimensional even though it really has a fifth dimension, Einstein and Bergmann describe a long thin tube
(fig. 3). 
 The idea is that if one looks up close, the tube is two-dimensional, since one can see its large length and its narrow width (and the tube  even appears three-dimensional if one can see the thickness of the material from it is made), while if one looks at it from a big distance, the tube appears one-dimensional as only its length is discernable.

  Previous authors had sometimes assumed that by fiat we are not allowed to observe the fifth dimension, or that it is part of the equations
but not part of physical reality.   Either way, this makes  the theory unsatisfying and unpredictive to modern eyes.  Instead, Einstein and Bergmann are very close to saying that although the equations of nature treat all five dimensions alike, the fifth dimension is much smaller in the world we live in and hence much harder to observe.  This is a very modern idea, and probably the closest Einstein came to an idea about how to unify the forces that is still important today.  Almost forty years later, Cremmer and Scherk [5] introduced this idea in so many words, and expressed it in the more modern language of spontaneous symmetry breaking (the broken symmetry being that between the fifth dimension, which is small, and the others, which are large).  Ever since, Kaluza-Klein theory has been an important ingredient in attempts to unify the forces.

\begin{figure}
\begin{center}
\includegraphics[width=3in]{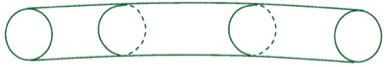}
\end{center}
\caption{To explain why a fifth dimension might be unobserved in everyday life, Einstein and Bergmann describe a long thin tube, as sketched
here.  (The drawing that actually appears in their paper shows instead a thin flat strip, which illustrates the same idea.)}
\end{figure}

However, Einstein drew back from the idea almost at once --  in the last part of the same paper -- and I believe never pursued it in his later writing.  
Instead, Einstein and Bergmann impose a constraint (on the existence of closed geodesics of specified length) which then leads them into what to a modern reader look like rather complicated calculations.  A subsequent paper [6] by Einstein, Bergmann, and V. Bargmann  is devoted to these calculations.

Einstein and Bergmann do not explain why they introduce the constraint that (in modern language) explicitly breaks the symmetry between the fifth dimension and the first four.  This constraint is sufficiently artificial-looking that there must have been a good reason for them to introduce it.  One obvious reason comes right to mind.  Had they not introduced their constraint, the fluctuations in the length of the fifth dimension would have given rise to a massless scalar field in four dimensions (sometimes called the radion or dilaton in modern treatments).  From a macroscopic point of view, the gravitational part of the theory, leaving electromagnetism aside, would not be pure General Relativity but rather a hybrid of General Relativity and the Nordstrom theory of gravity, the one that Einstein had rejected and over which he had triumphed.  Such a hybrid theory is nowadays often called a scalar-tensor theory (the scalar being the spin zero field on which Nordstrom based his early proposal for a relativistic
theory of gravity, while the tensor is Einstein's metric tensor).  

I first read the Einstein-Bergmann paper and some of the other early papers on Kaluza-Klein theory over 30 years ago, because I was curious to what extent the 
Cremmer-Scherk notion [5] of spontaneous compactification had antecedents in the old literature.  Clearly Einstein and Bergmann had come close, but their decision to introduce a constraint that spoiled the simplicity and beauty of the theory cried out for explanation.  The obvious explanation was that in 1938, Einstein simply did not want to modify General Relativity into a scalar-tensor theory.

I believe that my correspondence with Peter Bergmann (figs. 1 and 2) confirms  my interpretation.  He wrote, ``The reason for restricting the forms of the metric components $g_{5\dots}$  is explained in some detail on p. 275 of my relativity text.   Briefly the underlying reason is that this theory is not a scalar-tensor theory (Brans-Dicke type) but retains macroscopically $g_{55}=1$."  The text in question is ref. [6]; when Bergmann says that the theory ``is not" a scalar-tensor theory, I think he means that it is not desired or intended for it to be one. Bergmann went on to say that  imposing the constraint was a ``blemish'' that undermined the claim of the theory to being a unified theory (my comments are in brackets):  ``$\dots$  in the original K-K theory the five-dimensional symmetry is not broken `spontaneously,' but from the very beginning by the postulate of a Killing field.  And in the EBB [Einstein-Bergmann-Bargmann] modification it is broken by the topological postulate [the assumption that the fifth dimension is a circle], together with Einstein's geodesic postulate [the constraint that removes the scalar].  Either way, the break-up into the four dimensional symmetry plus a $\mathrm R$- or ${\mathrm S}^1$ gauge group is built into the theory prior to the consideration of any dynamic laws. This may well be considered a blemish of these theories, in that they are not really `unitary' [i.e., unified], contrary to the expectations of their originators.''  In the last remark, in my reading, Bergmann acknowledges that the
cost of removing the unwanted massless scalar was to compromise the beauty of the theory.

Perhaps a younger Einstein would have followed Kaluza-Klein theory to what today seems like the logical conclusion, and would have predicted a scalar-tensor modification of General Relativity. It would have been a bold prediction.  Einstein did not choose to go down that road.  As it turns out, the prediction would have failed the
test of later experiments, as we know now.  But it still would have been a memorable milestone, predicting a new field as a result of trying to unify the forces, and introducing questions that physicists attempting to unify the forces still grapple with today -- in  the moduli problem of string theory, for instance.  

According to a historical survey by Appelquist, Chodos, and Freund (see the introduction to [4]), it was left for Jordan and Thirry, a decade later, to describe the scalar-tensor theory that arises naturally from five-dimensional Kaluza-Klein theory, if no special constraint is imposed.  (Thirry's extremely short paper is reprinted in [4].)  Later, C. Brans and R. Dicke studied scalar-tensor theories systematically, and this gave part of the impetus for Dicke's classic experiments improving some of the tests of General Relativity.

In words that could inspire any physicist today, Einstein and Bergmann summarized their article: ``\dots It is much more satisfactory to introduce the fifth dimension not only formally, but to assign it some physical  meaning.  Nevertheless there is no contradiction with the empirical four dimensional character of physical space."  This they had explained well. But they also had written, ``Kaluza's aim was undoubtedly to obtain some new physical aspect for gravitation and electricity by introducing a unitary field structure.  This end was, however, not achieved."  Here one would have to say that Einstein and Bergmann evidently did not like the new physical aspect of gravitation and electricity that their reasoning must have suggested to them.

\bigskip
\bigskip
I thank P. Steinhardt for discussions.

\newpage
\noindent {\bf References}
\bigskip

\bigskip

\noindent [1] A. Einstein and P. Bergmann, ``On A Generalization Of Kaluza's Theory Of Electricity,"  {\it Annals of Mathematics} {\bf 39} (1938) p. 683.

\bigskip
\noindent [2] Th. Kaluza, ``On The Unification Problem In Physics," Sitzungsberichte  Pruss. Acad.  Sci.  (1921) p. 966, reprinted in English translation in reference [4].

\bigskip

\noindent [3]   O. Klein, ``Quantum Theory And Five-Dimensional Theory Of Relativity," Z.Phys. 37 (1926) p. 895, reprinted in English translation in reference [4]; ``The Atomicity Of Electricity As A Quantum Theory Law,'' {\it Nature} {\bf 118} (1926) p. 516, reprinted in [4].

\bigskip

\noindent [4] T. Appelquist, A. Chodos, and P. G. O. Freund, eds., {\it Modern Kaluza-Klein Theories}  (Addison-Wesley, 1987).

\bigskip

\noindent [5]  E. Cremmer and J. Scherk, ``Dual Models In Four Dimensions With Internal Symmetries," {\it Nucl. Phys.}  {\bf B103} (1976)  p. 399, ``Spontaneous Compactification Of Space In An Einstein Yang-Mills Higgs Theory," {\it Nucl. Phys.} {\bf B108} (1976) p. 409.

\bigskip

\noindent [6]  A. Einstein, V. Bargmann, and P. G. Bergmann, ``On The Five-Dimensional Representation Of Gravitation And Electricity,''  in Theodore von Karman Anniversary Volume (Cal Tech, 1941) p. 212.

\bigskip

\noindent [7] P. Bergmann, {\it An Introduction To The Theory Of Relativity,} (Prentice-Hall, 1942).

\end{document}